\documentclass[]{mn2e}
\usepackage{lscape,graphicx}

\newcommand{\hI}{\mbox{${\rm H\,I}$}}

\newcommand{\lya}{\mbox{${\rm Ly}\alpha$}}

\newcommand{\apg}{\:^{>}_{\sim}\:}
\newcommand{\apl}{\:^{<}_{\sim}\:}
\newcommand{\etal}{\ensuremath{\mbox{et~al.}}}

\providecommand{\kms}{\,\ensuremath{\rm{km\,s}^{-1}}}

\title[NIR Spectra of Two $z\apg 1.5$ GRB Hosts]{Near-infrared
  Spectroscopy of GRB Host Galaxies at $z\apg 1.5$: Insights 
  into Host Galaxy Dynamics and Interpretations of Afterglow
  Absorption Spectra} \author[H.-W. Chen]{Hsiao-Wen
  Chen$^{1}$\thanks{E-mail:
    hchen@oddjob.uchicago.edu} \\
  \\
  $^{1}$Dept.\ of Astronomy \& Astrophysics and Kavli Institute for
  Cosmological Physics, University of Chicago, Chicago, IL, 60637,
  U.S.A.}
\begin{document}



\maketitle

\label{firstpage}

\begin{abstract}

  This paper presents near-infrared echellette spectra of faint
  galaxies in the fields around GRB\,050820A at redshift $z=2.613$ and
  GRB\,060418 at $z=1.490$.  The spectroscopic data show that both
  GRBs originate in a dynamic environment of interacting galaxies
  separated by $< 15\ h^{-1}$ kpc in projected distance and
  $|\Delta\,v|\apl 60$ \kms\ in line-of-sight velocity.  The optical
  afterglows revealed in early-epoch Hubble Space Telescope images are
  at least $2.5\ h^{-1}$ kpc (or $0.4''$) away from the high surface
  brightness regions of the interacting members, indicating that the
  GRB events occurred either in the outskirts of a compact
  star-forming galaxy or in a low surface brightness satellite.
  Comparisons of the systemic redshifts of the host galaxies and the
  velocity distribution of absorbing clouds revealed in early-time
  afterglow spectra further show that the majority of the absorbing
  clouds are redshifted from these compact star-forming galaxies.
  These include the gas producing fine-structure absorption lines at
  physical distances $d\sim \mbox{a few}\times 100$ pc from the GRB
  afterglow.  The lack of blueshifted absorbing clouds and the spatial
  offset of the GRB event from the star-forming regions make it
  difficult to attribute the observed large velocity spread
  ($\sim\,200-400$ \kms) of absorbing gas in the GRB host to
  galactic-scale outflows.  We consider a scenario in which the GRB
  event occurred in a dwarf satellite of the interacting group and
  interpret the broad absorption signatures in the afterglow spectra
  as a collective effect of the turbulent halo gas and the host
  star-forming ISM.  We briefly discuss the implications for the
  absorption properties observed in the afterglow spectra.

\end{abstract}


\section{Introduction}

Optical afterglows of $\gamma$-ray bursts (GRBs) provide a novel
alternative for probing the interstellar medium, galactic halos, and
intergalactic gas in the distant universe (e.g.\ Vreeswijk \etal\
2004; Chen \etal\ 2005a; Prochaska, Chen, \& Bloom 2006; Vreeswijk
\etal\ 2007; Chen \etal\ 2007a; Fox \etal\ 2008; D'Elia \etal\ 2009;
Ledoux \etal\ 2009).  Because long-duration GRBs are believed to
originate in the catastrophic death of massive stars (e.g.\ Woosley \&
Bloom 2006) and because massive stars evolve rapidly, these GRBs also
provide a powerful internal light source for probing young, active
star-forming regions well into the epoch of re-ionization (e.g.\
Wijers \etal\ 1998; Tanvir \etal\ 2009; Salvaterra \etal\ 2009).
Unlike quasars, optical afterglows disappear after a few months and do
not hamper follow-up searches for GRB host galaxies and other faint
galaxies near the afterglow sightlines, allowing direct comparison
studies of ISM and stellar properties of distant galaxies.

Early-time afterglow spectra have revealed numerous absorption
features due to resonance and fine-structure transitions that enable
accurate measurements of the chemical composition, dust content, and
kinematics of both GRB host galaxies and intervening galaxies along
the lines of sight (e.g.\ Savaglio 2006; Prochaska \etal\ 2007).  In
particular, the absorption-line profiles have routinely shown complex
gas kinematics in the GRB host environments (e.g.\ Prochaska \etal\
2008a).  The physical origin of the observed large velocity spread is
unclear due to unknown emission properties of the host galaxies.  But
because of a massive star origin of GRB progenitors, a natural
explanation for the observed large velocity spread in the gas
foreground to the afterglows is starburst driven outflows.  However,
constraining the outflow velocity requires knowledge of the systemic
velocity of the host galaxy.

Here we present results from a pilot program to obtain spectroscopic
confirmation of GRB host galaxies at $z\apg 1.5$.  The primary
objective is to determine the systemic redshifts of the host galaxies
for studying gas flows in the GRB host environment.  At high redshift,
prominent ISM emission lines, such as [O\,II]3728, H$\beta$\,4862, and
H$\alpha$\,6564, are redshifted into the near-infrared wavelength
range at $\lambda>0.9\,\mu$m.  Near-infrared spectroscopy of faint
galaxies is challenging due to numerous strong atmospheric OH lines.
To increase the efficiency of the spectroscopic program, we utilize a
new near-infrared echellette spectrograph that offers sufficient
resolution over a broad wavelength range from $\lambda= 0.8\,\mu$m to
$\lambda=2.5 \,\mu$m for resolving galaxy emission features from OH
skylines.

We have selected galaxies in the fields around GRB\,050820A at
$z=2.613$ and GRB\,060418 at $z=1.490$ for the pilot spectroscopic
program.  For each GRB, multi-epoch optical images of the field are
available in the HST data archive that allow a precise and accurate
determination (better than $0.05''$) of the location of the optical
afterglow (e.g.\ Chen \etal\ 2009; Figure 1).  Candidate host galaxies
have been identified based on their proximity to the afterglow
position in the HST images.  In addition, high spectral resolution,
high $S/N$ early-time afterglow spectra are also available for the two
events permitting detailed characterizations of the absorption
properties of the host environment (Ellison \etal\ 2006; Prochaska
\etal\ 2007, 2008a; Vreeswijk \etal\ 2007).  We note that while
spectroscopic confirmations are already available for a handful of GRB
host galaxies at $z\apg 1.5$, including GRB\,971214 at $z=3.42$
(Kulkarni \etal\ 1998), GRB\,000926 at $z=2.04$ (Fynbo \etal\ 2002),
GRB\,011211 at $z=2.14$ (Fynbo \etal\ 2003), and GRB\,021004 at
$z=2.33$ (Jakobsson \etal\ 2005), these have been based on
observations of \lya\ emission and do not yield an accurate systemic
redshift of the host (e.g.\ Steidel \etal\ 2010).  Near-infrared
spectra of high-redshift GRB host galaxies have so far been available
only for the host of GRB\,021004 at $z=2.33$ (Castro-Tirado \etal\
2010).

This paper is organized as follows.  In Section 2, we describe the
spectroscopic observations of faint galaxies in the targeted GRB
fields and provide a summary of the data reduction and analysis
procedures.  In Section 3, we present the rest-frame optical spectra
of the GRB host galaxies.  In Section 4, we compare the systemic
velocities of the host galaxies with the velocity distribution of
gaseous clouds revealed in early-time afterglow spectra, and explore
scenarios that explain the observed complex gas kinematics in GRB host
environment.  Finally, we discuss the implications of the
spectroscopic observations in Section 5.  We adopt a flat $\Lambda$
cosmology, $\Omega_{\rm M}=0.3$ and $\Omega_\Lambda = 0.7$, with a
dimensionless Hubble constant $h = H_0/(100 \ {\rm km} \ {\rm s}^{-1}\
{\rm Mpc}^{-1})$ throughout the paper.

\begin{figure}
\includegraphics[scale=0.475]{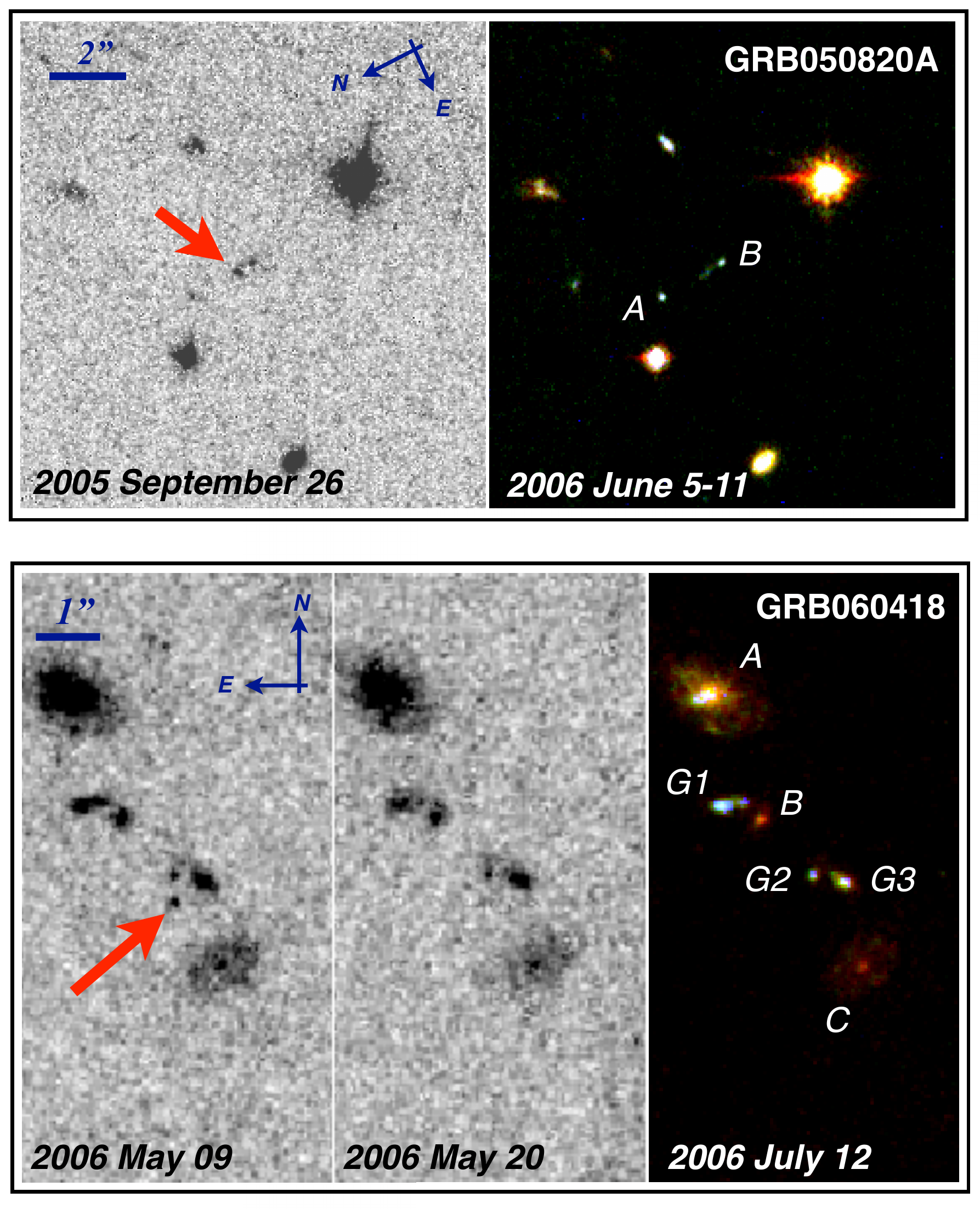}
\caption{Multi-epoch optical images of the field around GRB\,050820A
  (top) and the field around GRB\,060418 (bottom) obtained using the
  Hubble Space Telescope (HST) Advanced Camera for Surveys (ACS) and
  retrieved from the HST data archive (PID 10551; PI: S.\ Kulkarni).
  The early-epoch image of the field around GRB\,050820A was obtained
  using ACS and the F850LP fileter, and the late-epoch image was
  formed by combining stacks of images obtained using the F625W,
  F775W, and F850LP filters.  Galaxies $A$ and $B$ are discussed in
  \S\ 3.1 (see also Chen \etal\ 2009).  The early-epoch images of the
  field around GRB\,060418 were obtained using ACS and the F775W
  fileter, and the late-epoch image was formed by combining stacks of
  images obtained using the F555W, F625W, and F775W filters.  Galaxies
  $A$, $B$, $C$, $G1$, $G2$, and $G3$ are discussed in \S\ 3.2 (see
  also Pollack \etal\ 2009).  The optical transient (OT) in each field
  is apparent in the early-epoch images (marked by the arrow).  An
  extended, low surface brightness source is revealed at the location
  of GRB\,050820A in the late-epoch image.  No object is seen at the
  location of GRB\,060418 in the late-epoch image to a 5-$\sigma$
  limit of $AB({\rm F775W})=27.2$ over a $0.5''$ diameter aperture.}
\end{figure}

\section{Near-infrared Echellette Observations and Data Reduction}

We observed faint galaxies in the field around GRB\,050820A and
GRB\,060418 using the Folded port InfraRed Echellette (FIRE)
spectrograph (Simcoe et al.\ 2010) on the Magellan Baade Telescope.
FIRE offers high throughput and high spectral resolution (${\rm
  FWHM}\approx 50$ \kms) over a broad wavelength range from $\lambda=
0.8\,\mu$m to $\lambda=2.5 \,\mu$m in a single setup.  It is therefore
an efficient tool for studying faint galaxies at $z\apg 1$.  Here we
describe the observations and data reduction procedures.

\subsection{FIRE Observations}

The observations were carried out on the nights of 11 and 12 June 2011
using a $1''$ slit under mean seeing conditions of $0.5''$.  In
echellette mode, FIRE offers a slit length of $\approx 6''$ across the
spatial direction.  We were able to observe multiple galaxies in one
set up to optimize the spectroscopic observing efficiency.
For the field around GRB\,050820A, we were able to orient the slit to
include the low surface brightness host candidate of the GRB and the
nearby compact galaxies $A$ and $B$ at $\theta<1.5''$ from the
afterglow sightline (top panel of Figure 1; see also Chen \etal\
2009).  For the field around GRB\,060418, we were able to orient the
slit to include the group of galaxies that are likely the GRB host
(galaxies $G1$, $G2$, and $G3$ in the bottom panel of Figure 1) and
galaxy $B$ (see also Pollack \etal\ 2009) that are at $\theta\apl 2''$
from the afterglow sightline.

Exposures were taken in dithering ($ABBA$) mode in sets of four, each
of 900 s duration.  The separations between postions A and B were
about $2.5''$.  Calibration frames for wavelength solution were
obtained immediatly after each set of science exposures using an
internal ThAr lamp.  We also observed a nearby A0V star every hour for
calibrating the telluric features.  Flat-field calibration frames were
taken during the afternoon.  Additional twilight flat-field frames
were also obtained for correcting the non-uniform slit illumination
pattern.  A total exposure time of 3.5 hours was obtained for the
echellette spectroscopic observations of faint galaxies around
GRB\,050820A.  A total exposure time of 3 hours was obtained for the
group of galaxies around GRB\,060418.

\begin{figure*}
\includegraphics[scale=0.55]{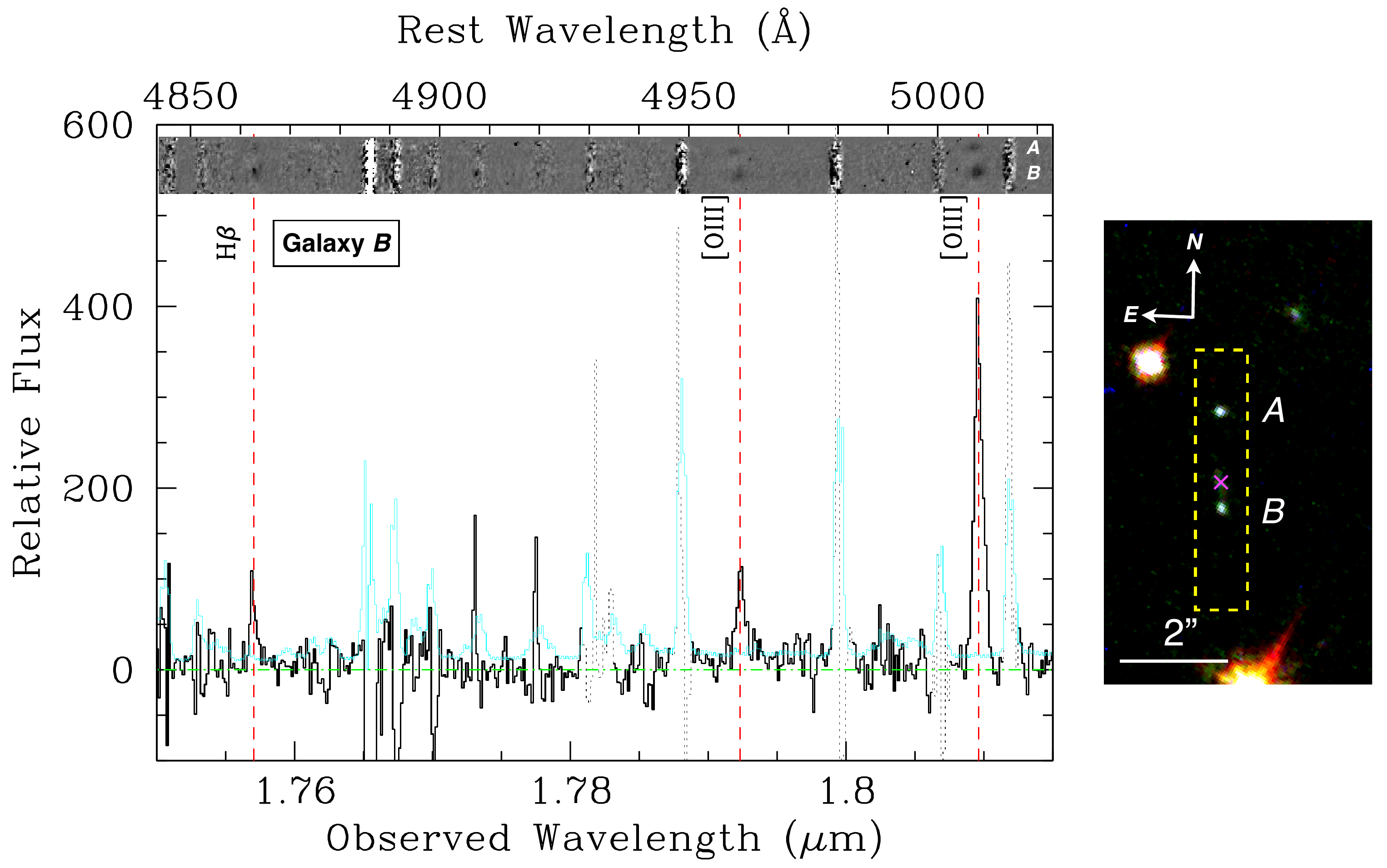}
\caption{Observations of galaxies along the sightline toward
  GRB\,050820A.  The right panel shows the image of the field from the
  top panel of Figure 1.  The image is rotated to orient the slit
  along the vertical direction.  The cross marks the position of the
  optical afterglow observed in early-epoch ACS images (Figure 1).  
  Galaxy A at angular
  distance $\theta \approx 1.3''$ and galaxy B at $\theta\approx
  0.4''$ were considered likely Mg\,II absorbing galaxies at $z=1.43$
  and $z=0.69$ (Chen \etal\ 2009).  The dashed rectangle shows roughly
  the slit coverage of the FIRE observations.  The left panel
  shows a portion of the stacked FIRE spectrum.  At the top, we
  show a rectified two-dimensional spectral image.  The OH sky lines
  have been subtracted, revealing two sets of emission-line features
  from galaxies $A$ (top) and $B$ (bottom).  These emission lines are
  consistent with H$\beta$\,4862, [O\,III]\,4960, and [O\,III]\,5008
  at $z= 2.6128$ and $z=2.6134$ for galaxies $A$ and $B$,
  respectively.  The one-dimensional spectrum at the bottom shows the
  emission features from galaxy $B$ as highlighted by vertical dashed
  lines.  Contaminating sky residuals are dotted out for clarity.  The
  1-$\sigma$ error spectrum is presented in the cyan histogram. }
\end{figure*}

\subsection{Data Reduction}

The FIRE spectral data were processed and reduced using a custom-built
reduction pipeline.  Data reduction and spectrum extraction of our
FIRE frames are challenging owing to two separate factors.  First, the
echellete spectra are distorted both in the dispersion and
cross-dispersion directions, and the spectral images contain numerous
tilted OH emission lines in the sky background.  Second, nearly all of
our targeted galaxies are faint and exhibit little/no trace of
continuum.  Therefore, accurate spectral traces cannot be
determined using the science exposures alone.

To overcome these difficulties, we modified a data reduction pipeline
that was originally developed by George Becker for processing optical
echellete spectral frames (see e.g.\ Chen \etal\ 2010).  This data
reduction pipeline utilizes a sky subtraction algorithm outlined in
Kelson (2003), which takes advantage of the fine-sampling of sky lines
afforded by tilted spectral images and determines an accurate
two-dimensional model of the sky spectrum to be removed from
individual frames.  The slit tilt was determined empirically based on
the locations of telluric standards along the cross-dispersion
direction.  A two-dimensional model of the wavelength solution in
vacuum units was then determined using available ThAr frames and
corrected for heliocentric motion.  Based on a comparison with OH sky
lines, we find that the accuracy of the wavelength solution is better
than 1 \AA.

To extract emission-line only spectra, the data reduction pipeline
first determines a trace solution across all echellette orders using
the telluric standard frames obtained close in time and location to
the targeted galaxies.  An offset along the slit is then applied to
match the trace to the emission-line features seen in the
two-dimensional sky-subtracted frames.  The observed A0V telluric
standard were also useful for calibrating the sensitivity functions of
individual echellette orders.  We determined the throughput function
for each echellette order by calibrating the observed telluric
standard to a model A0V spectrum.

To optimize the accuracy of sky-subtraction and the signal-to-noise of
the final extracted spectra, we first performed sky subtraction in
individual spectral images.  Next, we registered the sky-subtracted
science frames using emission-line features visible in the data and
formed a stacked two-dimensional spectral image for each targeted
galaxy.  Individual echellette orders were then extracted from the
stacked frame using a boxcar extraction routine, and calibrated using
the sensitivity function derived from the observed telluric standard.
A final spectrum of the galaxy was formed by coadding these
flux-calibrated orders.

\section{Spectral properties of Individual GRB Host Galaxies}

The FIRE observations provided a spectroscopic confirmation of the
host galaxies of two GRBs at $z\apg 1.5$.  Here we present the
rest-frame optical spectra of the hosts of GRB\,050820A and
GRB\,060418 at $z=2.613$ and $z=1.49$, respectively, and compare the
systemic velocities of the hosts with the velocity field of gaseous
clouds revealed in early-time afterglow spectra.

\begin{figure}
\includegraphics[scale=0.48]{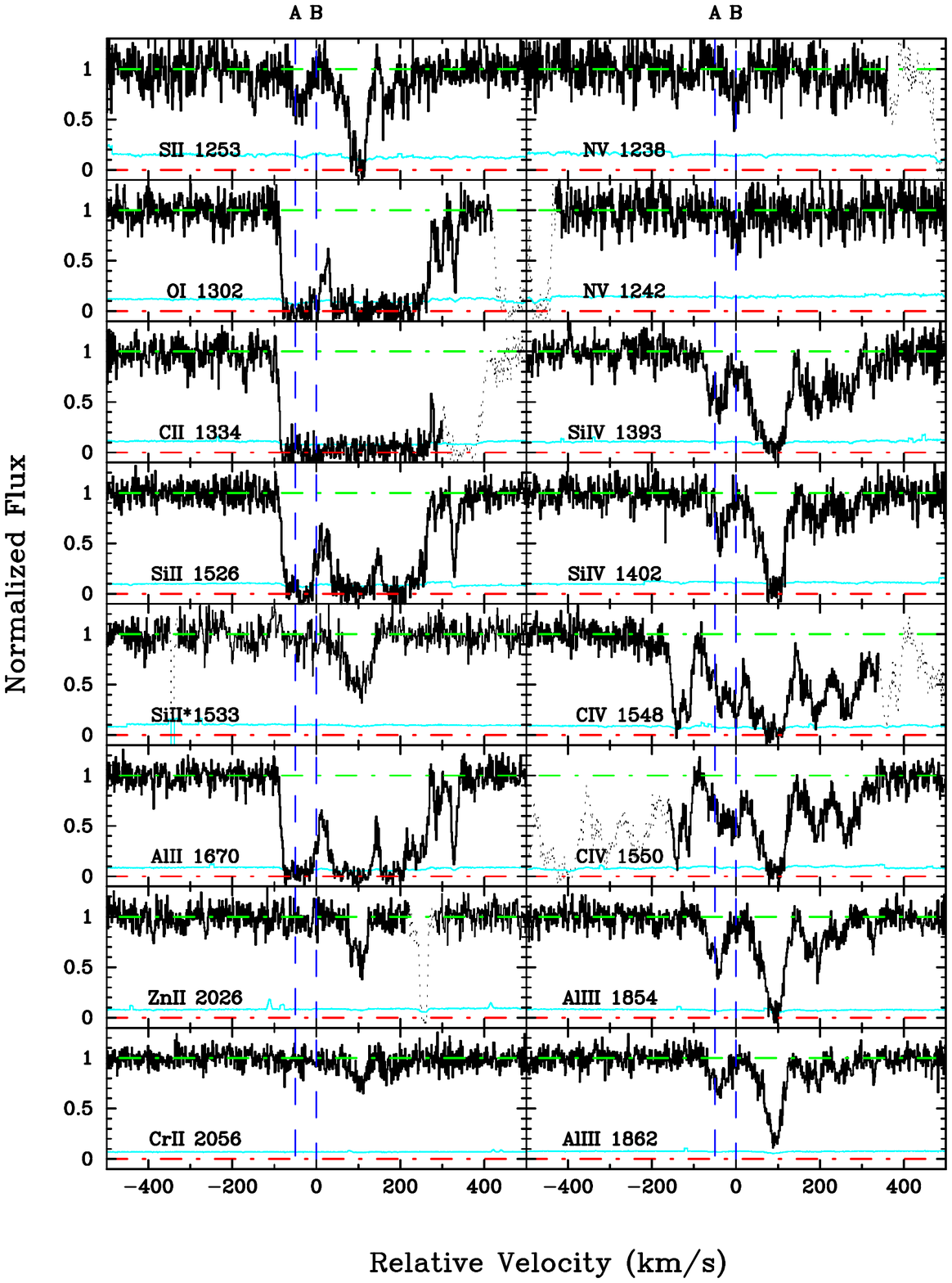}
\caption{A subsample of the absorption profiles of low- and
  high-ionization species associated with the host of GRB\,050820A
  from early-time afterglow spectra (Prochaska \etal\ 2007).
  Contaminating features are dotted out for clarity.  Zero velocity
  corresponds to $z=2.6134$, the redshift of galaxy $B$ at
  $\theta=0.44''$ (corresponding to a projected distance of $\rho=2.5\
  h^{-1}$ kpc) from the GRB sightline in Figure 2.  Galaxy $A$ at
  $\theta=1.34''$ (corresponding to a projected distance of $\rho=7.5\
  h^{-1}$ kpc) is found to offset from the systemic redshift of galaxy
  $B$ by $\Delta\,v=-50$ \kms.  This GRB host exhibits complex
  line-of-sight gas kinematics with multiple absorbing components
  spreading over a velocity range from $\Delta\,v=-140$ \kms\ to
  $\Delta\,v=+320$ \kms.  The dominant neutral gas component occurs at
  $\Delta\,v=+108$ \kms, as indicated by the low-ionization resonance
  and fine-structure transitions Zn\,II\,2026, Cr\,II\,2056, and
  Si\,II*\,1533.  In contrast, the highly ionized gaseous clump that
  produces the N\,V\,1238, 1242 doublets appears to coincide in
  velocity space with galaxy $B$.}
\end{figure}

\subsection{The Host of GRB\,050820A at $z=2.613$}

GRB\,050820A was detected by {\it Swift} (Page \etal\ 2005).  An
optical transient (OT) was identified less than 1 hr after the trigger
(Fox \& Cenko 2005; Vestrand \etal\ 2006).  High-resolution
($\delta\,v \approx 7-10$ \kms) echelle spectra of the afterglow
obtained shortly after the burst (Ledoux \etal\ 2005; Prochaska \etal\
2007; Fox \etal\ 2008) allowed accurate measurements of the source
redshift and chemical abudances in the ISM of the host galaxy.  The
GRB was found to occur at $z=2.6147$.  The host galaxy was found to
contain a total neutral hydrogen column density of
$\log\,N(\hI)=21.0\pm 0.1$ and metallicity $[{\rm S}/{\rm H}]=
-0.63\pm 0.11$ along the GRB sightline (Prochaska \etal\ 2007).  The
observed relative abundances $[{\rm S}/{\rm Fe}]= +0.97\pm 0.09$
suggests a dust-to-gas ratio in the host ISM comparable to what is
seen in the Small Magellanic Cloud with a visual extinction of
$A_V\approx 0.08$.  No trace of H$_2$ was found despite the large
$N(\hI)$ and moderate metallicity.  The molecular fraction of the host
ISM was constrained to be $f_{\rm H_2}\equiv 2N({\rm
  H_2})/[N(\hI)+2\,N({\rm H_2})] < 10^{-6.5}$ (Tumlinson \etal\ 2007).
Finally, two strong Mg\,II absorbers were found at $z=0.692$ and
$z=1.430$ with $W(2796)=2.99\pm 0.03$ \AA\ and $W(2796)=1.9\pm 0.1$
\AA\ in the rest frame, respectively (Prochaska \etal\ 2007).

Photometric properties of faint galaxies in the field around
GRB\,050820A have been studied by Chen \etal\ (2009) based on an
analysis of available multi-epoch ACS images of the field in the HST
data archive.  These authors identified an extended low surface
brightness galaxy at the location of the OT in late-time images after
the OT disappeared (top panel of Figure 1).  In addition, they
identified two compact sources at $\theta< 1.5''$ of the GRB sightline
($A$ and $B$ in the top panel of Figure 1).  All three objects have
comparable observed optical magnitudes of $AB({\rm F775W})\approx
26.2$ (Chen \etal\ 2009), despite the large difference in the apparent
surface brightness.  Chen \etal\ (2009) considered the low surface
brightness object as the GRB host candidate, given its coincident
position with the OT, and galaxies $A$ and $B$ as likely Mg\,II
absorbing galaxies in the foreground.

\subsubsection{Rest-frame Optical Spectra of  the Host Candidates}

We present in the left panel of Figure 2 a portion of the stacked FIRE
spectrum.  The rectified two-dimensional spectral image over the
observed wavelength range of $\lambda=1.75-1.815\,\mu$m is shown at
the top and the extracted one-dimensional spectrum of galaxy $B$ is
shown at the bottom.  The slit size and orientation of the FIRE
observations (dashed rectangle in the right panel of Figure 2) allowed
us to observe the candidate GRB host and galaxies $A$ and $B$ in a
single set-up.  We detect two sets of emission-line features that are
offset in both spatial and spectral directions in the stacked FIRE
spectral image (left panel of Figure 2).  No trace of continuum
emission is visible.  The spatial separation of the emission lines is
consistent with the spatial separation of galaxies $A$ and $B$,
although the low surface brightness candidate host and galaxy $B$ are
expected to be blended in ground-based data due to seeing.

The spectral features of galaxies $A$ and $B$ appear to be very
similar.  The emission lines are consistent with H$\beta$\,4862,
[O\,III]\,4960, and [O\,III]\,5008 at $z= 2.6128\pm 0.0002$ and
$z=2.6134\pm 0.0002$ for galaxies $A$ and $B$, respectively.  At this
redshift, we also detect H$\alpha$ at $\lambda\approx 23,712$ \AA\
(not shown in Figure 2) but not [O\,II]\,3727, 3729 in the FIRE
spectra.  The known $H$-band brightness limit, $AB(H)>26$, from Chen
\etal\ (2009) constrains the rest-frame absolute $B$-band magnitude to
be fainter than $M_{AB}(B)-5\,\log\,h=-18.5$ for both galaxies $A$ and
$B$.  The angular separation between galaxies $A$ and $B$,
$\theta=1.78''$, corresponds to a physical projected distance of
$\rho_{\rm AB}=10\,h^{-1}$ kpc.  The line-of-sight velocity offset
between the two objects is $\Delta\,v=50\pm 8$ \kms.  Such small
separations in projected distance and velocity space indicate that
galaxies $A$ and $B$ are an interacting pair associated with the host
of GRB\,050820A at $z=2.613$, and that GRB\,050820A occurred in either
one of the galaxies or in a low surface brightness satellite.

\subsubsection{The Velocity Field of Absorbing Clouds Along the GRB Sightline}

The FIRE observations have revealed interacting galaxies ($A$, $B$,
and possibly the extended low surface brightness feature in the top
panel of Figure 1) in the vicinity of GRB\,050820A.  Here we compare
the spatially and spectrally resolved galactic dynamics with the
velocity distribution of gaseous clouds revealed in early-time
afterglow spectra.

In Figure 3, we present a subsample of the absorption profiles of low-
and high-ionization species associated with the host of GRB\,050820A
from early-time afterglow spectra (Prochaska \etal\ 2007).  Zero
velocity corresponds to $z=2.6134$, the redshift of galaxy $B$ at
$\theta=0.44''$ (corresponding to a projected distance of $\rho=2.5\
h^{-1}$ kpc) from the GRB sightline in Figure 2.  Galaxy $A$ at
$\theta=1.34''$ (corresponding to a projected distance of $\rho=7.5\
h^{-1}$ kpc) is found to offset from the systemic redshift of galaxy
$B$ by $\Delta\,v=-50$ \kms.  

The absorption profiles of the gas foreground to GRB\,050820A are
characterized by a large number of components, covering a velocity
range from $\Delta\,v=-140$ \kms\ to $\Delta\,v=+320$ \kms\ of the
host galaxies.  The majority of the absorbing clouds appear to be
redshifted from the systemic velocities of both members of the
interacting system, including the dominant neutral gas component at
$\Delta\,v=+108$ \kms.
In contrast, the highly ionized gaseous clump that produces the
N\,V\,1238, 1242 doublets appears to coincide in velocity space with
galaxy $B$.

\begin{figure*}
\includegraphics[scale=0.55]{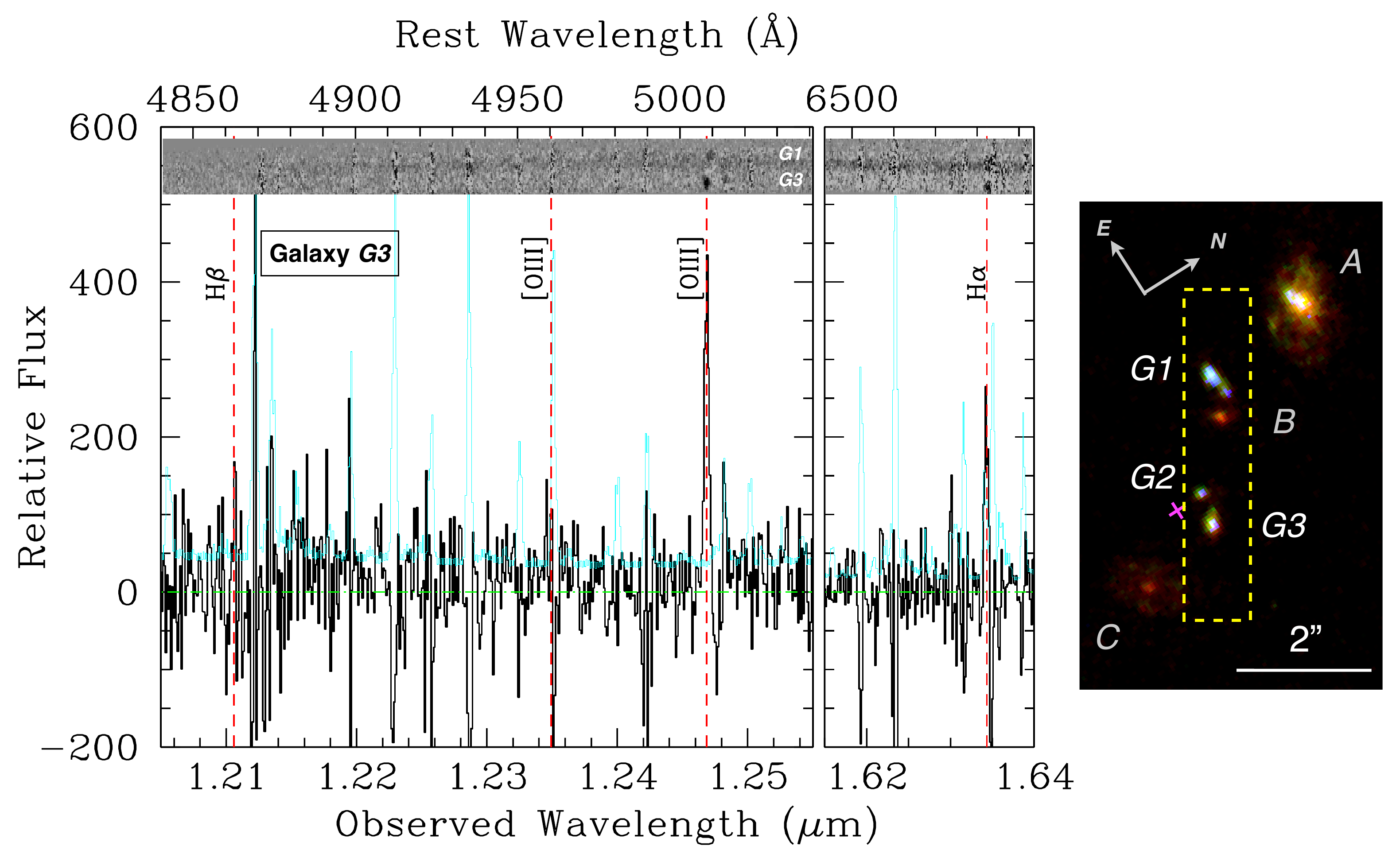}
\caption{Observations of galaxies along the sightline toward
  GRB\,060418.  The right panel shows the image of the field from the
  bottom panel of Figure 1.  The image is rotated to orient the slit
  along the vertical direction.  The cross marks the position of the
  optical afterglow observed in early-epoch ACS images (Figure 1).
  The slit size and orientation of the FIRE observations (dashed
  rectangle) allowed us to simultaneously observe the galaxies in the
  interacting group, as well as galaxy $B$.  The left panel shows
  portions of the stacked FIRE spectrum.  At the top, we show 
  rectified two-dimensional spectral images.  At the bottom, we present
  the one-dimensional spectrum of galaxy $G3$ with emission features
  highlighted by vertical dashed lines and the associated 1-$\sigma$
  error spectrum shown in the cyan histogram.  The OH sky lines have
  been subtracted in both frames.  The 2D sky-subtracted spectral
  image reveals a faint trace of continuum emission and two sets of
  two emission-line features.  The emission-line features are
  spatially offset from the continuum.  We interpret the faint trace
  of continuum spectrum as due to galaxy $B$ and the emission lines as
  due to galaxies $G1$ (top) and $G3$ (bottom).  These emission lines
  are consistent with [O\,III]\,5008 and H$\alpha$ at $z= 1.4901$ and
  $z=1.4896$ for galaxies $G1$ and $G3$, respectively.  Near these
  redshifts, the [O\,III]\,4960 emission line is blended with a strong
  OH sky line.  The spectrum of galaxy $B$ does not show any clear
  line features for measuring its redshift.}
\end{figure*}

\subsection{The Host of GRB\,060418 at $z=1.490$}

GRB\,060418 was detected by {\it Swift} (Falcone \etal\ 2006).  The OT
was identified nearly instantaneously after the trigger (Falcone
\etal\ 2006).  Echelle spectra of the afterglow obtained shortly after
the burst revealed numerous low-ion resonant and fine-structure
transitions that established the redshift of the GRB at $z=1.491$
(Ellison \etal\ 2006; Prochaska \etal\ 2007; Vreeswijk \etal\ 2007).
At this redshift, however, the host HI\,1215 transition is not covered
in ground-based spectra.  The neutral hydrogen column density of the
host is therefore unknown.  The presence of abundant heavy ions, e.g.\
$N({\rm Si\,II})>15.89$ (Prochaska \etal\ 2007), suggests that the
host ISM contains a neutral hydrogen column density of
$\log\,N(\hI)>20.3$ for solar metallicity.  The $N(\hI)$ limit would
be still higher, if the gas contains sub-solar metallicity.  The
observed relative abundance $[{\rm Cr}/{\rm Zn}]= -0.31\pm 0.06$
(Prochaska \etal\ 2007) also suggests a mild dust depletion (e.g.\
Pettini \etal\ 1997).  Finally, this GRB sightline exhibits an
extraordinarily high density of strong Mg\,II absorbers over a small
redshift range.  In addition to the host, three Mg\,II absorbers were
found at $z=0.603$, 0.656, and 1.107 with $W(2796)=1.27\pm 0.01$ \AA,
$0.97\pm 0.01$ \AA, and $1.84\pm 0.02$ \AA\ in the rest frame,
respectively (Prochaska \etal\ 2007).

Faint galaxies in the field around GRB\,060418 have been studied by
Pollack \etal\ (2009) based on an analysis of available multi-epoch
ground-based and space-based broad-band images of the field.  These
authors identified an unusually high surface density of faint galaxies
at $\theta\apl 3.5''$ from the GRB sightline in available ACS images
from the HST data archive (Figure 1).  This high surface density of
faint galaxies appears to correspond to the large number of strong
Mg\,II absorbers revealed in early-time afterglow spectra.  However,
no object is found at the location of the OT (Figure 1).  Given the
consistent colors and possible interacting morphologies, Pollack
\etal\ (2009) speculated that galaxies $G1$, $G2$, and $G3$ are an
interacting galaxy group hosting the GRB and attributed galaxies $A$,
$B$, and $C$ to the strong Mg\,II absorbers at lower redshifts.

Pollack \etal\ (2009) attempted optical spectroscopy of these faint
galaxies and confirmed that galaxy $A$ is indeed associated with the
Mg\,II absorber at $z=0.656$.  However, galaxies $B$ and $C$ remain
unconfirmed due to a lack of spectral features in the optical data.

\subsubsection{Rest-frame Optical Spectra of  the Host Candidates}

We present in the left panel of Figure 4 portions of the stacked FIRE
spectrum.  Rectified two-dimensional spectral images over the observed
wavelength ranges of $\lambda=1.205-1.225\,\mu$m and
$\lambda=1.615-1.640\,\mu$m are shown at the top and the extracted
one-dimensional spectrum of galaxy $G3$ is shown at the bottom.  The
slit size and orientation of the FIRE observations (dashed rectangle
in the right panel of Figure 4) allowed simultaneous observations of
the candidate GRB host, $G1$, $G2$, and $G3$, and galaxy $B$ in a
single set-up.  Galaxies $G2$ and $G3$ are expected to be blended in
the FIRE observations due to seeing.  Similar to the field of
GRB\,050820A, we also detect two sets of emission-line features for
the galaxies around GRB\,060418 that are offset in both spatial and
spectral directions from each other in the stacked FIRE spectral
image.

Figure 4 also shows a faint trace of continuum emission immediately
adjacent to the emission features at the top of the spectral frame.  A
careful examination of the sky-subtracted, two-dimensional spectral
image indicates that the continuum spectrum is spatially offset from
the emission-line features.  The spatial separation of the emission
lines is consistent with the spatial separation of galaxies $G1$ and
$G2$.  We therefore interpret the continuum spectrum as due to galaxy
$B$ and the emission lines as due to galaxies $G1$ and $G3$.  Due to a
lack of spectral features, we are unable to determine the redshift of
galaxy $B$.  The emission lines of $G1$ and $G3$ are consistent with
[O\,III]\,5008 and H$\alpha$ at $z= 1.4901\pm 0.0002$ and $z=1.4896\pm
0.0002$, respectively.  Near these redshifts, the [O\,III]\,4960
emission line is blended with a strong OH sky line.  The observed
$H$-band brightness limit, $AB(H)>25$, from Pollack \etal\ (2009)
constrains the rest-frame absolute $B$-band magnitude to be fainter
than $M_{AB}(B)-5\,\log\,h=-18.4$ for galaxy $G3$.  The angular
separation between galaxies $G1$ and $G3$, $\theta=2.15''$,
corresponds to $\rho_{\rm G13}=12.7\,h^{-1}$ kpc.  The line-of-sight
velocity offset between the two objects is $\Delta\,v=60\pm 8$ \kms.
Similar to the configuration of the host of GRB\,050820A, such small
separations in projected distance and velocity space indicate that
galaxies $G1$ and $G3$ (and likely $G2$) are an interacting group,
hosting GRB\,060418 at $z=1.490$.

\subsubsection{The Velocity Field of Absorbing Clouds Along the GRB Sightline}

Similar to the field around GRB\,050820A, the FIRE observations have
revealed interacting galaxies in the vicinity of GRB\,060418 ($G1$ and
$G3$ in the bottom panel of Figure 1).  Here we compare the spatially
and spectrally resolved galactic dynamics with the velocity
distribution of gaseous clouds revealed in early-time afterglow
spectra.

In Figure 5, we present a subsample of the absorption profiles of low-
and high-ionization species associated with the host of GRB\,060418
from early-time afterglow spectra (Prochaska \etal\ 2007).  Zero
velocity corresponds to $z=1.4896$, the redshift of galaxy $G3$ at
$\theta=0.53''$ (corresponding to a projected distance of $\rho=3.1\
h^{-1}$ kpc) from the GRB sightline in Figure 3.  Galaxy $G1$ at
$\theta=2.01''$ (corresponding to a projected distance of $\rho=11.9\
h^{-1}$ kpc) is found offset from the systemic redshift of galaxy $G3$
by $\Delta\,v=+60$ \kms.

The absorption profiles of the gas foreground to GRB\,060418 are
characterized by a classic 'edge-leading' signature, with the dominant
neutral gas components occuring at $\Delta\,v\approx +60$ \kms\ and a
blueshifted tail (most prominent in the Mg\,II and C\,IV absorption
transitions) extending to $\Delta\,v=-140$ \kms\ from the systemic
velocity of galaxy $G3$.  In this case, the dominant absorbing
components as indicated by the low-ionization resonance and
fine-structure transitions Ni\,II 1741 and Fe\,II\,2389 appear to
coincide in velocity space with galaxy $G1$ at nearly $4\times$ the
projected distance of $G3$ from the afterglow sightline.

An edge-leading absorption profile is often considered a signature of
an underlying rotating disk (e.g.\ Lanzetta \& Bowen 1992).  Recall
that the GRB afterglow is an internal source that probes only part of
the host ISM.  The matching velocity between $G1$ and the strongest
absorption component makes it difficult to explain the gas kinematics
based on a rotation disk around $G1$.  If instead the absorption
signature is due to a rotating disk around $G3$, then redshifted
strong components suggests that the GRB occured in the back side of
the disk.  However, such ordered motion appears to be inconsistent
with the irregular morphologies of galaxies $G1$, $G2$, and $G3$
(Figure 4).

\begin{figure}
\includegraphics[scale=0.45]{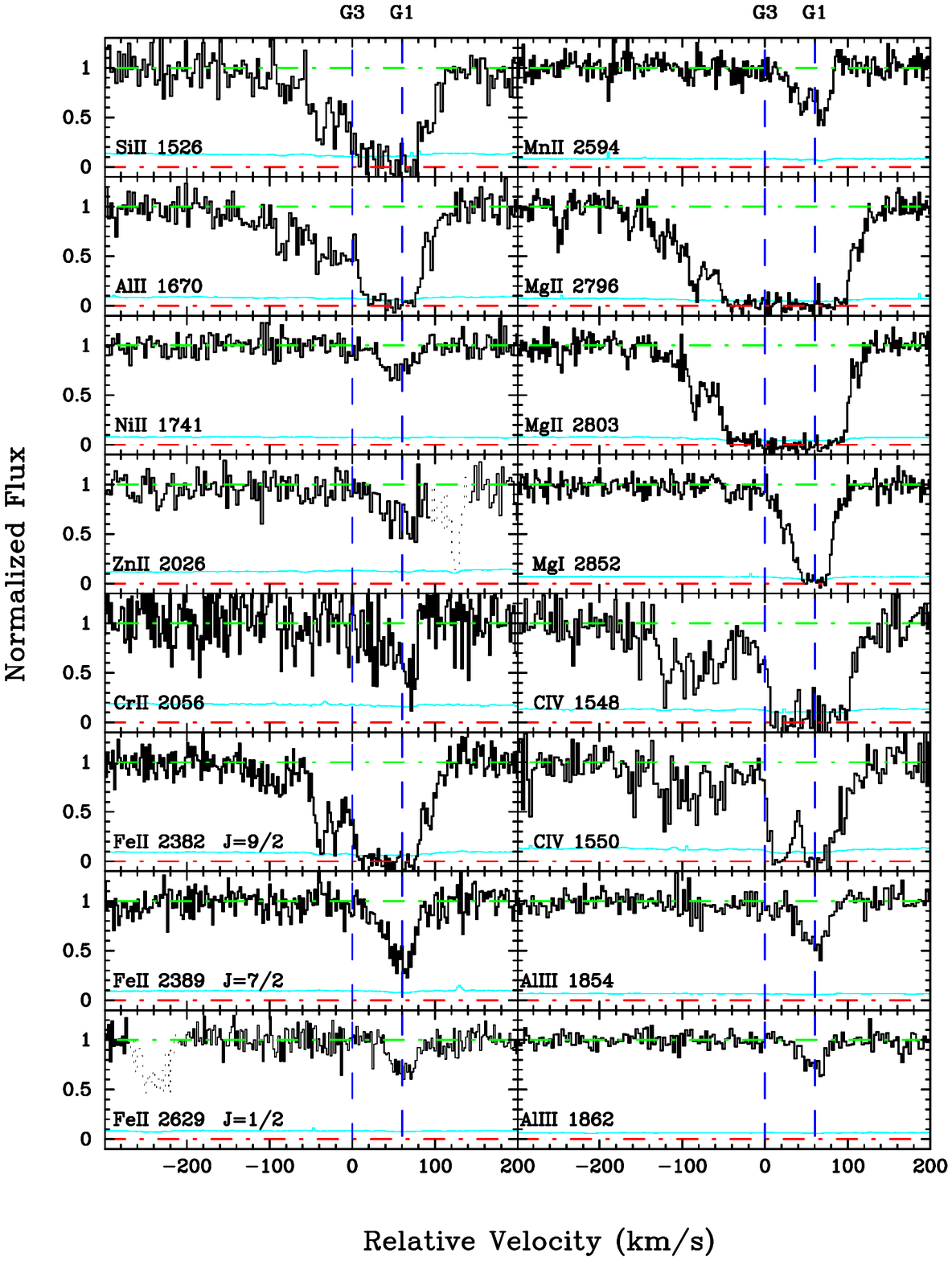}
\caption{A subsample of the absorption profiles of low- and
  high-ionization species associated with the host of GRB\,060418 from
  early-time afterglow spectra (Prochaska \etal\ 2007).  Contaminating
  features are dotted out for clarity.  Zero velocity corresponds to
  $z=1.4896$, the redshift of galaxy $G3$ at $\theta=0.53''$
  (corresponding to a projected distance of $\rho=3.1\ h^{-1}$ kpc)
  from the GRB sightline in Figure 3.  Galaxy $G1$ at $\theta=2.01''$
  (corresponding to a projected distance of $\rho=11.9\ h^{-1}$ kpc)
  is found to offset from the systemic redshift of galaxy $G3$ by
  $\Delta\,v=+60$ \kms.  This GRB host also exhibits complex
  line-of-sight gas kinematics with multiple components spreading over
  a velocity range from $\Delta\,v=-140$ \kms\ to $\Delta\,v=+120$
  \kms.  The dominant absorbing components occur at $\Delta\,v=+60$
  \kms, as indicated by the low-ionization resonance and
  fine-structure transitions Ni\,II 1741 and Fe\,II\,2389.  These
  low-ionization species appear to coincide in velocity space with
  galaxy $G1$.}
\end{figure}

\section{Dynamics of GRB Host Galaxies}

Combining the systemic velocities of the host star-forming regions and
the line-of-sight velocity distributions of gaseous clouds in front of
the GRB afterglows has allowed us to establish an absolute velocity
field for the host galaxies of GRB\,050820A and GRB\,060418.  The
observed velocity field together with known ISM and stellar properties
of the host galaxies offers important insights into the galactic
environment of GRB hosts and the nature of the complex gas kinematics
revealed in afterglow absorption spectra.

A unique aspect in afterglow absorption-line studies is the presence
of excited ions in the host ISM as a result of UV pumping by the GRB
afterglow (e.g.\ Prochaska, Chen, \& Bloom 2006; Dessauges-Zavadsky
\etal\ 2006; Vreeswijk \etal\ 2007, 2011).  Given known radiation
field from light-curve observations of the optical transient and the
observed abundances of these excited ions, we have constrained the
physical distance of the dominant neutral absorbing component, such as
the component at $\Delta\,v=+108$ \kms\ for GRB\,050820A in Figure 3
and the components at $\Delta\,v\approx +60$ \kms\ for GRB\,060418 in
Figure 5, at $d\sim \mbox{a few}\times 100$ pc.  This distance scale
places these neutral gaseous clouds outside of the birth cloud of the
GRB progenitor but still within the same star-forming region where the
progenitor star formed.

In the case of GRB\,050820A, the majority of the absorbing clouds,
including the gas at $d\sim \mbox{a few}\times 100$ pc from the GRB
afterglow, are observed to be redshifted from both galaxies $A$ and
$B$ at projected distances $\rho=2.5-7.5\ h^{-1}$ kpc away.  In the
case of GRB\,060418, the majority of the absorbing clouds are also
found to be redshifted from galaxy $G3$ at merely $\rho=3.1\ h^{-1}$
kpc away.  The neutral gaseous clouds at $d\sim \mbox{a few}\times
100$ pc from the GRB progenitor coincide with the systemic velocity of
$G1$ at $\rho=11.9\ h^{-1}$ kpc away, nearly $4\times$ the projected
distance of $G3$.

In both cases, the GRB did not occur in the brightest regions of their
host galaxies (cf.\ Fruchter \etal\ 2006).  Instead, they were either
in the outskirts of a compact star-forming galaxy or in a low surface
brightness satellite.  If they were in the outskirts of one of the
interacting galaxies, then the lack of blueshifted absorbing
components relative to the systemic velocities of the galaxies makes
it difficult to attribute the observed complex gas kinematics along
the afterglow sightline to galactic-scale outflows from the
interacting galaxies.  We note that although blueshifted absorbing
components are seen relative to $G3$ along the GRB\,060418 sightline,
the HST images clearly show a spatial offset between the GRB event and
the interacting members $G2$ and $G3$ (Figure 1).  Different from the
self-absorption seen in distant star-forming galaxies (e.g.\ Weiner
\etal\ 2009), the GRB sightline does not probe directly into the
compact star-forming regions.  While outflows from galaxies $G1$ and
$G3$ may contribute to some of the absorption components at the blue
end, outflows alone is difficult to explain the 'edge-leading'
absorption signatures along the afterglow sightline (Figure 5).

To understand the origin of the observed gas kinematics, we continue
the discussion assuming that the GRBs occurred in a nearby dwarf
satellite of the spectrocopically identified interacting galaxies.  We
focus on GRB\,050820A as an example 
but the same scenario can be applied to explain the observations
of GRB\,060418.  We illustrate the relative alignment in the cartoon
shown in Figure 6.

We consider scenarios in which the interacting galaxies are either in
the background or foreground of the host star-forming region.  If the
GRB arises in a satellite ($S$ shown in grey in Figure 6) in front of
the interacting galaxies, then the observed redshifted motion of the
host ISM relative to galaxies $A$ and $B$ suggests that the host
satellite galaxy is falling toward the interacting galaxies.  In this
scenario, the GRB sightline does not probe the ISM of either galaxy
$A$ or $B$.  Therefore the large velocity spread of $\sim 450$ \kms\
observed in the afterglow spectrum can only be explained by possible
turbulence in the local star-forming ISM (including supernova driven
outflows, if present) adjacent to birth site of the GRB progenitor and
by halo gas.

If the host star-forming region is behind the interacting galaxies,
then the observed redshifted motion of the host ISM suggests that the
GRB arises in a tidal dwarf galaxy ($S$ shown in dark in Figure 6)
escaping galaxy $B$ as galaxies $A$ and $B$ interact with each other.
A local example of such system is Arp 245 (Duc \etal\ 2000).  The
large velocity spread of absorbing clouds in the afterglow spectrum
can therefore be accounted for by a collective velocity field of the
gas local to the star-forming region where the GRB progenitor formed,
the ISM of the interacting galaxies, tidal debris, and accreted halo
gas.  The lack of absorption systems blueshifted from galaxies $A$ and
$B$ along the GRB sightline also constrains the extent of outflows at
$\rho<2.5\ (7.5)\ h^{-1}$ kpc from galaxy $B$ ($A$).

\begin{figure}
\begin{center}
\includegraphics[scale=0.5]{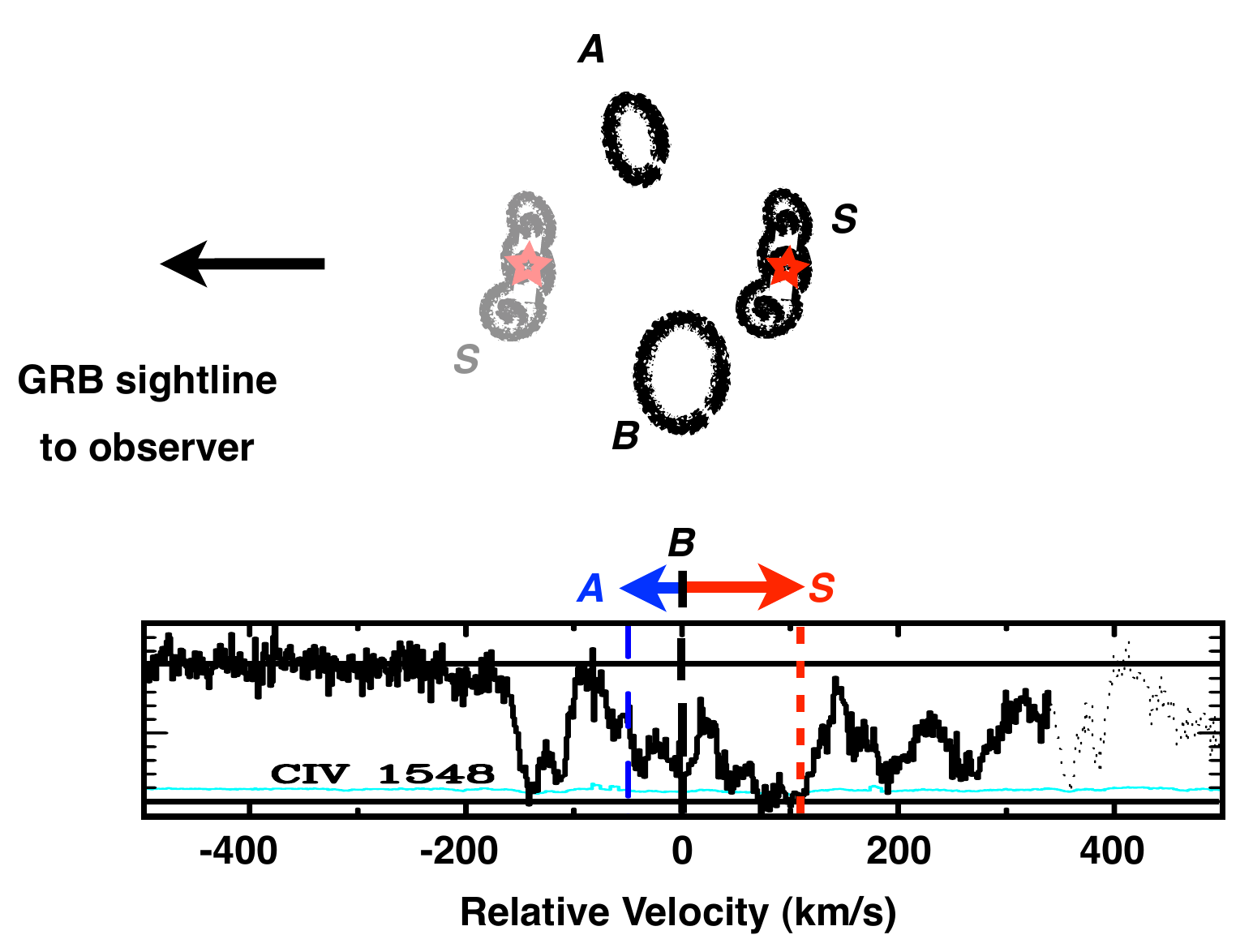}
\caption{Cartoon illustrating possible configurations between the host
  star-forming satellite of GRB\,050820A and the interacting galaxies
  $A$ and $B$ to explain the observed line-of-sight velocity
  distribution of the absorbing gas shown in the bottom panel.  If the
  GRB arises in a satellite ($S$ shown in grey) in front of the
  interacting galaxies, then the observed redshifted motion of the
  host ISM relative to galaxies $A$ and $B$ suggests that the host
  satellite galaxy is falling toward the interacting galaxies.  In
  this scenario, the GRB sightline does not probe the ISM of either
  galaxy $A$ or $B$, but only the local star-forming ISM of the GRB
  progenitor and halo gas.  If the host star-forming region is behind
  the interacting galaxies, then the observed redshifted motion of the
  host ISM suggests that the GRB arises in a tidal dwarf galaxy ($S$
  shown in dark) escaping galaxy $B$ as galaxies $A$ and $B$ interact
  with each other.  In this scenario, the star-forming region local to
  the GRB progenitor, the ISM of the interacting galaxies, tidal
  debris, and accreted halo gas are all expected to contribute to the
  absorption features in the afterglow spectra.}
\end{center}
\end{figure}

\section{Discussion}

Near-infrared echellette spectra presented here have allowed us to go
beyond a spectroscopic confirmation of two $z\apg 1.5$ GRB host
galaxies and obtain an accurate and precise measurement of the
systemic redshifts of the host galaxies for studying gas flows in the
GRB host environment.  We find that the hosts of GRB\,050820A and
GRB\,060418 share three common features.

First, both GRBs are found to originate in a dynamic environment of
interacting galaxies separated by $10-12\ h^{-1}$ kpc in projected
distances and $|\Delta\,v|\apl 60$ \kms\ in line-of-sight velocity
separation.  We have interpreted the high surface brightness
star-forming regions as merging galaxies based primarily on the
optical morphologies (Figure 1).  While it is also possible that these
compact sources are star-forming regions within the same galaxy, we
consider this an unlikely scenario given the observed morphologies and
the projected separations of $\sim 2''$ or $\sim 10\ h^{-1}$ kpc.
Previous studies based on deep HST images have shown that star-forming
galaxies at $z=2-3$ have on-average half-light radii of $\approx 0.2"$
(e.g.\ Bouwens \etal\ 2004) or $\approx 1.1\ h^{-1}$ kpc (e.g.\
Ralfeski \etal\ 2011; Law \etal\ 2011).  Under a single object
scenario, the projected distances observed between $A$ and $B$ around
GRB\,050820A or between $G1$ and $G3$ around GRB\,060418 would imply a
galaxy that is at least five times the typical size of $z\sim 2$
galaxies.

The optical morphologies of the host galaxies of GRB\,050820A and
GRB\,060418 resemble the hosts of GRB\,000926 at $z=2.04$ and
GRB\,011211 at $z=2.14$, displaying a similarly disturbed morphology
of interacting objects.  Together they show that a large fraction
(4/10) of known GRB host galaxies at $z\apg 1.5$ (Kulkarni \etal\
1998; Jensen \etal\ 2001; Fynbo \etal\ 2002; Fynbo \etal\ 2005;
Vreeswijk \etal\ 2004, 2006; Chen \etal\ 2009, 2010) may arise in
interacting galaxies.  Superstar clusters are commonly seen in
interacting galaxies (e.g.\ de Grijs \etal\ 2002), which serve as a
nature birth place for the progenitors of GRBs (e.g.\ Chen \etal\
2007b).

Second, the majority of the absorbing gaseous clouds observed in the
early-time afterglow spectra, including the gas producing
fine-structure absorption lines at $d\sim \mbox{a few}\times 100$ pc
from the GRB afterglow, are found to be redshifted from the
interacting galaxies.  This is different from what is seen in the host
of GRB\,021004 at $z=2.33$, for which all of the absorbing gaseous
clouds observed in the early-time afterglow spectra are found to be
blueshifted from the systemic velocity of the host galaxy over the
range of $|\Delta\,v|=100-800$ \kms.  At low redshifts, such kinematic
study is known only for the host of GRB\,030329 at $z=0.16867$, which
shows absorption components primarily blueshifted from the systemic
velocity up to $|\Delta\,v|=200$ \kms\ (Th\"one \etal\ 2007).  

Recall that GRB afterglows serve as an internal light of the host
galaxies.  The predominantly redshifted motion found for the gas in
front of GRB\,050820A and GRB\,060418 therefore suggests inflows.
Likewise, the observed blueshifted motion in GRB\,021004 and
GRB\,030329 suggests outflows.
The mixture of blueshifted and redshifted gas on velocity scales
greater than 200 \kms\ in front of the GRB progenitor site underscores
the complex and turbulent gas inflows and outflows around distant
young star-forming galaxies (e.g.\ Haehnelt \etal\ 1998; Agertz \etal\
2009).  This is different from a more organized rotational motion seen
in some damped \lya\ absorbing galaxies (Chen \etal\ 2005b).

Finally, GRB\,050820A and GRB\,060418 did not occur in the brightest
star-forming regions of their host galaxies like GRB\,000926,
GRB\,011211, or other long-duration bursts (cf.\ Fruchter \etal\
2006).  In contrast, the closest, high surface brightness star-forming
region is found to be at least $\rho=2.5\ h^{-1}$ kpc (or $0.4''$)
away, roughly three times the median projected offset found for
long-duration GRBs (e.g.\ Bloom \etal\ 2002).  This spatial offset and
the lack of blueshifted absorbing systems relatively to the
star-forming members suggest that the GRB event occurred in a low
surface brightness tidal tail or satellite.  If GRB\,050820A arises in
a tidal tail of galaxy $B$ as galaxies $A$ and $B$ approach each other
(Figure 6), then the observed strength and velocity offset of the
N\,V\,1238, 1242 absorption doublets can be naturally explained by the
conduction interface of stripped gas (e.g.\ Indebetouw \& Shull 2004)
from the interacting galaxies (cf.\ Prochaska \etal\ 2008b).

In addition, satellite galaxies or tidal tails contain relatively more
pristine gas around otherwise chemically evolved star-forming galaxies
(e.g.\ Thilker \etal\ 2009).  The broad range of galactic environement
probed by GRB events, from the luminous host example of GRB\,021004 to
possible faint dwarf satellites for the hosts of GRB\,050820 and
GRB\,060418, is qualitatively consistent with the large scatter
observed in the ISM metallicity of GRB host galaxies. (e.g.\ Savaglio
2010).  The large velocity spread and low metallicity expected in
tidal tails may also explain the large scatters seen in the
velocity--metallicity correlation (Prochaska \etal\ 2008a; Ledoux
\etal\ 2006).

In summary, we show that comparisons of the systemic redshifts of the
host galaxies and the velocity distribution of absorbing clouds
revealed in early-time afterglow spectra provide new insights into the
nature of GRB host galaxies and their environments.  Galactic-scale
winds are not the only factor that drives the observed large velocity
spread of gaseous clouds along GRB sightlines.  Four of 10 GRB hosts
known at $z\apg 1.5$ exhibit similar morphologies showing interacting
galaxies.  Galaxy interactions can be effective in producing
chemically enriched materials away from more concentrated star-forming
regions found in deep galaxy surveys.

Lastly, we also note that the absorbing galaxies responsible for the
two strong Mg\,II absorbers at $z=0.692$ ($W(2796)=2.99\pm 0.03$ \AA)
and $z=1.430$ ($W(2796)=1.9\pm 0.1$ \AA) remain missing around the
sightline toward GRB\,050820A.  No galaxies brighter than $AB({\rm
  F775W})=27.5$ (over a $0.5''$ diameter aperture) are seen at
$\theta<3.5''$ of the afterglow sightline in late-time HST ACS images.
Either the absorbing galaxies are extremely faint $<0.03 L_*$, or we
are seeing a large collection of absorbing clouds at projected
distance $\rho \apg 20\ h^{-1}$ kpc from a star-forming galaxy.

\section*{Acknowledgments}

I thank an anonymous referee for a swift review and helpful comments.
I also thank George Becker, Jean-Ren\'e Gauthier, Nick Gnedin, Andrey
Kravtsov, Lynn Matthews, Michael Rauch, and Rob Simcoe for helpful
discussions and comments.  This work was supported in part by NASA
through Grant HST-GO-12005.02 from the Space Telescope Science
Institute, which is operated by the Association of Universities for
Research in Astronomy, Inc., under NASA contract NAS 5-26555.

\label{lastpage}

\end{document}